\documentclass[options]{JHEP3} 

\usepackage{epsfig,multicol,psfig,graphicx,psfrag,amsthm,latexsym,multibox,amssymb,amsfonts}

\newcommand\fverb{\setbox\pippobox=\hbox\bgroup\verb}
\newcommand\fverbdo{\egroup\medskip\noindent%
			\fbox{\unhbox\pippobox}\ }
\newcommand\fverbit{\egroup\item[\fbox{\unhbox\pippobox}]}
\newbox\pippobox

\newcommand{\be}{\begin{equation}}
\newcommand{\ee}{\end{equation}}
\newcommand{\ba}{\begin{eqnarray}}
\newcommand{\ea}{\end{eqnarray}}

\csname @addtoreset\endcsname{equation}{section}

\title{The continuous spin limit of higher spin field equations}

\author{Xavier Bekaert\\\email{bekaert@ihes.fr}\\\\ Institut des Hautes \'Etudes Scientifiques, Le Bois-Marie\\
35 route de Chartres, 91440 Bures-sur-Yvette (France)}

\author{Jihad Mourad\\\email{mourad@th.u-psud.fr}\\\\APC \thanks{Unit\'e Mixte
de Recherche du CNRS (UMR 7164).}, Universit\'e  Paris VII,\\
2 place Jussieu, 75251 Paris Cedex 05 (France)\\\\
LPT \thanks{Unit\'e Mixte de Recherche du CNRS (UMR
8627).}, B\^at. 210 , Universit\'e Paris XI, \\ 91405 Orsay Cedex
(France)}

\preprint{\hepth{0509092}}

\abstract{We show that the Wigner equations describing the continuous spin
representations can be obtained as a  limit of massive higher-spin field
equations. The limit involves  a suitable scaling of the wave
function, the mass going to zero  and the spin to infinity
with their product being fixed. The result allows to transform the
Wigner equations to a gauge invariant Fronsdal-like form. We also
give the generalisation of the Wigner equations to higher
dimensions with fields belonging to  arbitrary
 representations of the massless little
group.}


\begin{document}

\section{Introduction}

The unitary irreducible representations (UIR) of the Poincar\'e
group $P_D=ISO(D-1,1)$, as Wigner has shown, are determined
by those of the little group \cite{Wigner:1939cj}. On the one
hand, the little group for massless particles in
$D$-dimensional Minkowski spacetime ${\mathbb R}^{D-1,1}$ is the
non-compact Euclidean group $E_{D-2}=ISO(D-2)$. Its UIR
are infinite-dimensional\footnote{ For spacetime dimension
$D\geqslant 4$.} except for the case where all the
translation-like generators vanish. The latter case characterises
the ``helicity"
representations whose little group is effectively
$SO(D-2)$. The generic case gives rise to the so called
``continuous spin" \cite{wi,Wigner:1963in,chak,Abbott:1976bb,Brink:2002zx}
representations\footnote{ They are
also called ``infinite
spin" representations \cite{Wigner:1963in}.} with a  nonvanishing
value of the second Casimir operator
$W=\mu^2\,$, where $\mu$ is a real parameter with the
dimension of a mass. Wigner proposed a set of
manifestly covariant equations to describe fields carrying these
UIR in four spacetime dimensions \cite{wi}. The wave function
depends on the usual spacetime coordinates and in addition on an
internal four-vector. These equations are reviewed in the second
section. Three of the Wigner equations allow to constrain this
four-vector to a transverse angle variable which is at the origin
of the ``continuous spin" name.

The massive representations, on the other hand, are determined by
representations of the rotation group $SO(D-1)$. From the group
theoretical point of view, the UIR of the  orthogonal and
Euclidean groups  are related by  an
In\"{o}n\"{u}-Wigner contraction $SO(D-1)\rightarrow E_{D-2}$
\cite{Inonu:1953sp}. It follows that one can obtain the continuous
spin representations from the massive ones in a massless
limit $m\rightarrow 0$. The second Casimir
operator is related to the spin $s$ of the particle as
$W=m^2s(s+D-3)$. In order to keep $W$ nonvanishing, the  massless limit must be such that
the product $s\,m$ remains  finite, $sm\rightarrow \mu\,$, so
that the spin goes to infinity, for a group theoretical discussion see
\cite{Khan:2004nj}.
The main goal of this paper is to  obtain
covariant wave equations for the continuous spin representations, in
any spacetime dimension, from massive higher-spin equations.

A massive spin-$s$ particle can be described by a rank-$s$ tensor
field \cite{pf} or, more conveniently, as a Kaluza-Klein mode of a massless
spin-$s$ particle \cite{Fronsdal:1978rb}
on a higher dimensional spacetime\footnote{{
The helicity little group in $D+1$ dimensions is the group
$SO(D-1)$, which is identified with the massive little group in
$D$ dimensions.}} with a nonvanishing momentum along the extra
dimension \cite{Singh,Aragone,Hallowell:2005np}. This gives rise, as reviewed in Section
\ref{massivehs}, to a collection of totally symmetric tensors
having ranks less than or equal to $s$. If one introduces an
auxiliary vector $u^\mu$, one can interpret the tensors as the
Taylor coefficients of the expansion in powers of $u$. We will
show, in Section \ref{frommhs}, that this is the way that the
Wigner internal vector arises and the Wigner equations emerge in
the contraction limit. This limit involves a proper rescaling
of the  wave function and the auxiliary variables in order to be
well behaved.
 Starting from the equations
of motion of the higher-spin particle in de Donder's gauge we get
the Wigner equations in the aforementioned limit. This
suggests that the Wigner equations correspond to a gauge-fixed
version of gauge invariant equations of motion. We show that this
is the case in Section \ref{fromFronsdal}, where we determine the
new gauge symmetries.
The equations can be formulated with a
restricted gauge invariance, in analogy with the Fronsdal
equations \cite{Fronsdal:1978rb}, or with an unconstrained gauge
parameter. In the second case one has to introduce a
``compensator" field \cite{compensator}. In Sections
\ref{generalis} and \ref{mixed} we discuss generalisations of the
equations we found to arbitrary UIR of $SO(D-3)$,
the little group of $E_{D-2}$, which we shall
call the ``short" little group \cite{Brink:2002zx}.
We shall consider the
spinorial  representation in Section \ref{generalis} and the ``exotic" 
ones in Section
\ref{mixed}. We collect our conclusions in Section 8.

There are many physical motivations to study higher-spin fields
and their limits. We refer, for example, to \cite{Sorokin:2004ie}
 for a comprehensive discussion.
Let us mention that String theory gives rise to particles with arbitrary
spins and a proper understanding of  its symmetries may
be approached by revealing the symmetries underlying the higher-spin fields and the constraints from the consistency of their interactions.
In this respect, some higher derivative string theories \cite{Savvidy:2003fx}
give rise to particles belonging to the continuous spin
representation. The way that conventional string theory and
this tensionless higher derivative one are related
may be clarified by the relation between the continuous spin fields
and the higher-spin massive fields which is the subject of this paper.

\section{The continuous spin representation}

Consider a massless  particle in $D$-dimensional spacetime\footnote{{
Our conventions are as follows: Greek indices such as $\mu$,
$\nu$, $\ldots$ denote spacetime indices running from $0$ to
$D-1$ while Latin indices such as $i$, $j$, $\ldots$ denote
transverse indices running from $1$ to $D-2$. The Minkowski metric
is mostly plus and reads in light-cone coordinates:
$ds^2=-2dx^+dx^-+dx^idx_i$. Dots denote contraction of (implicit)
spacetime indices.}}.
Let its momentum have zero components except for $p^+$
($V^{\pm}={1 \over \sqrt{2}}(V^0\pm V^{D-1}),
 V^+=-V_-$). The little group leaving the momentum invariant is
 generated by $M_{ij}$ and $M_{+i}=\pi_i$, they
 verify the Lie algebra
 \ba
 \left[\pi_i,\pi_j\right] &=& 0\,,\ \left[M_{ij},M_{kl}\right]=
 i(\delta_{jk}M_{il}
 -\delta_{ik}M_{jl}-\delta_{jl}M_{ik}+\delta_{il}M_{jk})\,,
 \nonumber\\
 \left[\pi_i,M_{kl}\right]&=&i(\delta_{ik}\pi_l-\delta_{il}\pi_k)\,,
 \ea
 which is the Lie algebra of the
 $D-2$ dimensional Euclidean group
 $E_{D-2}$. The Casimir $\pi^i\pi_i=\mu^2$ classifies the
 representations of $E_{D-2}$. If $\mu$ vanishes then the
 irreducible representations are given by those of $SO(D-2)$,
 these are the helicity states.
 When $\mu$ is nonvanishing we get the
 continuous spin representations. They are of infinite dimension
 and are determined by
 the UIR of the subgroup leaving a given
 $\pi^ i$ invariant, the short little group, $SO(D-3)$.

 In fact, the second Casimir operator of the Poincar\'e group,
 is given by
 \be
 W=-{1\over 2}p^2M_{\mu\nu}M^{\mu\nu}+M_{\mu\alpha}p^\alpha M^{\mu\beta}p_\beta.
 \ee
 It reduces to $\mu^2$ for the massless particle and to
 $m^2s(s+D-3)$ for the massive one. Here $s$ corresponds to the rank of the
 traceless completely symmetric tensor in $D$ spacetime dimensions.

\subsection{Wigner's wave equation}

A wave equation whose physical content is the single valued
continuous spin representation was proposed by Wigner \cite{wi}.
The wave function depends on two vectors: the momentum $p$
 and the additional vector $\xi$ which is dimensionless. There are two independent
equations obeyed by the wave function $\overline\Psi(p\,,\xi)$ and
two other which are consequences of these.
They read
\ba {\cal E}_1\overline\Psi\equiv p\cdot{\partial
\overline\Psi\over
\partial \xi}
-i\mu\, \overline\Psi&=&0,\label{epsilon1}\\
{\cal E}_2\overline\Psi\equiv (\xi^2-1)\overline\Psi&=&0\,. \label{epsilon2}\ea
The
first compatibility condition reads \be [{\cal E}_1,{\cal
E}_2]\overline\Psi\ \equiv \ 2\,{\cal E}_3\overline\Psi\ =\, 2\,
p\cdot\xi\,\overline\Psi=0,
\label{epsilon3}\ee and the second compatibility condition \be
[{\cal E}_1,{\cal E}_3]\overline\Psi \ \equiv\ {\cal E}_4\overline\Psi\ = \
p^2\,\overline\Psi=0\label{epsilon4} \ee
 is the mass-shell
constraint. There are no more compatibility conditions.
 These equations can be obtained as the first class constraints arising from
  a higher derivative classical action \cite{Zoller:1991hs}.

The equation (\ref{epsilon1}) reflects the fact that
the couples $(p\,,\xi)$ and $(p\,,\xi+\alpha p)$ are
physically equivalent for arbitrary $\alpha\in\mathbb R$. Indeed,
one gets
\be \overline\Psi(p\,,\xi+\alpha p)\,=\,e^{i\alpha
\mu}\,\overline\Psi(p\,,\xi) \label{jau}
\ee from Equation
(\ref{epsilon1}). The equation (\ref{epsilon2}) states that the
internal vector $\xi$ is a unit space-like vector while the
mass-shell condition (\ref{epsilon4}) states that the momentum is
light-like. From the equation (\ref{epsilon3}), one obtains that
the internal vector is transverse to the momentum. All together,
one finds that $\xi$ lives on the unit hypersphere $S^{D-3}$ of
the transverse hyperplane ${\mathbb R}^{D-2}$.
In brief, the
``continuous spin" degrees of freedom essentially correspond to
$D-3$ angular variables, whose Fourier conjugates are
discrete variables analogous to the usual spin degrees of
freedom.

\subsection{The Fourier transformed wave equation}

In fact, it is useful for later purposes to write the
equations (\ref{epsilon1}) - (\ref{epsilon4}) in terms of
$w\,$, the Fourier conjugate to $\xi\,$. The
equations  now read
\ba
(p\cdot w+\mu)\,\Psi=0\,,&&\label{cspin1}\\
\left({\partial \over \partial w}\cdot {\partial \over \partial
w}+1\right)\Psi=0\,,\label{cspin2}&&\\
\left(p\cdot{\partial \over \partial w}\right)\Psi=0\,,\label{cspin3}&&\\
p^2\,\Psi=0\,.&& \label{cspin4} \ea

In order to explicit the physical content of the equations, let us
consider a plane wave, \be \Psi(p,w)\,=\,\delta(p-p_0)\,\psi_{p_0}(w),
\ee with $p_0^2=0$. Suppose that the only non-vanishing
component of $p_0^\mu$ is $p_0^+$, then Equation (\ref{cspin1}) implies that
\be 
\psi_{p_0}(w)\,=\,\delta(w^-p_0^+-\mu)\,\,\phi\,(w^+,w^i), \ee where
$w^i$ are the transverse coordinates. Equation (\ref{cspin3})
implies that $\phi$ does not depend on $w^+$ and, finally,
Equation (\ref{cspin2}) becomes the Helmholtz equation
\be
{\partial^2 \phi\over
\partial w^i\partial w_i} +\phi=0. \label{tracecond}\ee

There are several formal ways to write the solutions to the
equation (\ref{tracecond}). A first way is to expand $\phi(w^i)$
in powers of $w$:
\be \phi(w^i)=\sum\limits_{n=0}^\infty{1\over
n!}\,\phi_{i_1\dots i_r}w^{i_1}\dots w^{i_n},\label{firstexp} \ee
with symmetric coefficients obeying\be \phi_{i_1\dots
i_n}=-\phi^{j}{}_{j\,i_1\dots i_n}.\label{trred} \ee

A second way is to to expand in spherical harmonics
as
\be
\phi(w^i)=\sum_{n=0}^{\infty}{f_n(r)\over n!}f_{i_1\dots i_n}
\hat w^{i_1}\dots\hat w^{i_n},\label{harm}
\ee
where $r^2=w_iw^i$, and $\hat w^i=w^i/r$ are the coordinates on the sphere
$S^{D-3}$.
In equation (\ref{harm}), the constant tensors $f_{i_1\dots i_n}$
are totally symmetric and traceless. The
function $f_n$ 
verifies the differential equation
\be
f''_n+{{D-3}\over r}f_n'+\left(1-{n(n+D-4)\over r^2}\right)f_n=0,\label{equat}
\ee
which results from Equation (\ref{tracecond}).
 The solution to the equation
(\ref{equat}) which is regular
at $r=0$
is given by
\be
f_n(r)=r^{2-{D\over 2}}J_{n+{D\over 2}-2}(r),
\ee
where $J_\nu$ is the Bessel function of the first kind.
Notice that each term in the expansion (\ref{harm}) is by itself a solution to
the Helmoltz equation. This was not the case of the first expansion.

In both expansions,  one gets totally symmetric tensors that
one is tempted to compare with the fields appearing in the
description of a massive higher-spin particle. The first
expansion (\ref{firstexp}) turns out to be the one which will
allow to make contact with the massive case.  Rougly speaking,
the point is that the physical components of a spin-$s$ massive
symmetric field correspond to a spin-$s$ irreducible
representation of the massive little group $SO(D-1)$, {\it i.e.} a
rank-$s$ traceless symmetric $D-1$ tensor $\phi_{I_1\ldots I_s}$
($I_k=1,\ldots,D-1$), which decomposes as a tower of totally
symmetric $D-2$ tensor $\phi_{i_1\ldots i_r}$ of rank $r$ running
from zero to $s$ and satisfying precisely (\ref{trred}). The
second expansion (\ref{harm}) has the merit of exhibiting the
physical content of the Wigner equations: the general
solution is given by  a sum of plane waves with functions over
the (internal) hypersphere $S^{D-3}$ as coefficients. We stress that the
continuous spin wave function $\Psi$ has a number of ``components"
which is infinite but {\it countable}. For a nonvanishing $\mu$, a Lorentz
transformation not belonging to the little group mixes the tensors of different
ranks. The mixing disappears when $\mu$ vanishes and in this limit
we get an infinite sum over
all helicity states represented above by the tensors $f_{i_1\dots i_n}$.

The Hilbert space of functions on $S^{D-3}$ carries the UIR
of the massless little group $E_{D-2}$ with a trivial representation of the
short little group $SO(D-3)$. We shall consider cases with
arbitrary irreducible representations of the latter in Section
\ref{generalis}.

\section{Massless and massive higher-spin fields}\label{massivehs}

A convenient way of obtaining the equations of motion for the
massive  higher-spin fields is to start from a massless spin-$s$
field in one extra space dimension and to compactify on a circle
with a non vanishing momentum.

\subsection{Massless higher-spin field}

The Fronsdal equation for a massless higher-spin field described
by a rank-$s$ totally symmetric tensor
$\varphi_{\mu_1\dots\mu_s}$ with a vanishing double trace is given
by \cite{Fronsdal:1978rb} \be
p^2\varphi_{\mu_1\dots\mu_s}-p_{(\mu_1}p^\nu
\varphi_{\mu_2\dots\mu_s)\nu}
+p_{(\mu_1}p_{\mu_2}\varphi_{\mu_3\dots\mu_s)\nu}{}^\nu=0,
\label{Fronsdalequ}\ee where the curly bracket denotes
complete symmetrisation  by summing over all
different permutations. These equations are invariant under the
gauge transformations 
\be \delta
\varphi_{\mu_1\dots\mu_s}=p_{(\mu_1}\varepsilon_{\mu_2\dots\mu_s)},\label{gtransf}
\ee
where the tensor $\varepsilon$ is traceless. If one introduces
an auxiliary  vector $u$ and defines \be \varphi(x,u)\,=\,{1\over
s!}\,\varphi_{\mu_1\dots\,\mu_s}\,u^{\mu_1}\dots u^{\mu_s},
\label{hdef}\ee then the Fronsdal equation (\ref{Fronsdalequ})
may be rewritten as \be \left[p^2-(p\cdot u)\left(p\cdot
{\partial \over
\partial u}\right)+{1\over 2}\,(p\cdot u)^2 \left({\partial \over \partial
u}\cdot{\partial \over \partial u}\right)\right]\varphi(x,u)=0\,.
\ee
The double trace constraint on the gauge field becomes
\be \left({\partial \over
\partial u}\cdot{\partial \over \partial u}\right)^2\varphi=0\,, \ee
and
the homogeneity in $u$ of the function $\varphi$ defined by
(\ref{hdef}) implies \be \left(u\cdot{\partial \over
\partial u}-s\right)\varphi=0\,. \ee
The gauge transformations
(\ref{gtransf}) read
\be
\delta \varphi\,=\,(p\cdot u)\, \varepsilon, \ee with \be
\left(u\cdot{\partial \over
\partial u}-s+1\right)\varepsilon=0 \ee and the tracelessness of
$\varepsilon$ leads to \be \left({\partial \over
\partial u}\cdot{\partial \over \partial u}\right)\,\varepsilon=0. \ee

In de Donder's gauge \be \left[p\cdot{\partial \over
\partial u}-{1 \over 2}\,(p\cdot u) \left({\partial \over \partial
u}\cdot{\partial \over
\partial u}\right)\right]\varphi=0, \ee the Fronsdal equation
(\ref{Fronsdalequ}) simplifies to $p^2\varphi=0$. There is a
residual gauge invariance with $\varepsilon$ subject, in addition
to the tracelessness condition, to \be p^2\,\varepsilon=0\,,\quad
\left(p\cdot{\partial \over
\partial u}\right)\,\varepsilon=0\,. \ee
The remaining gauge invariance \cite{m}
allows to
impose the tracelessness of $\varphi$ \be \left({\partial \over
\partial u}\cdot{\partial \over \partial u}\right)\, \varphi=0, \ee so
that de Donder's gauge becomes \be \left(p\cdot{\partial \over
\partial u}\right) \varphi=0, \ee
which expresses the transversality
of $\varphi$.
We can now define a gauge invariant field by \be
\Phi=\delta(u\cdot p) \,\varphi\,. \ee
The new field $\Phi$ is no
more a polynomial in $u$ but it is nevertheless a homogeneous
distribution in $u$ of degree $s-1$. It allows to write compatible and
covariant
equations describing the massless higher-spin degrees of freedom
\ba
(p\cdot u)\,\Phi=0,\label{physmassl}\\
\left({\partial \over \partial u}\cdot{\partial \over \partial
u}\right)\Phi=0,\\
\left(p\cdot {\partial \over \partial u}\right) \Phi=0,\\
p^2\,\Phi=0\,,\\
\left(u\cdot{\partial \over \partial
u}-s+1\right)\Phi=0\,.\label{spins}
 \ea
 In fact, Equations (\ref{physmassl}) - (\ref{spins})
 describe by themselves a massless rank-s totally symmetric field,
 it is no longer necessary to assume the polynomiality in $u$.
  Notice the similarity with the Wigner
equations, which have the same form except for constants in the
first and second equations and the absence of an analogue of the
last equation.

\subsection{Massive higher-spin field}

In order to get the equations for a massive particle we start with
a massless higher-spin particle in dimension $D+1$ and consider a
mode with $p^D=m$. We divide the $D+1$ auxiliary vector into
a $D$ vector $u$ and a scalar $v$ with respect to $SO(1,D-1)$.

{The Fronsdal equation (\ref{Fronsdalequ}) becomes
 \be \left[p^2+m^2-(p\cdot u+mv)\left(p\cdot
{\partial \over \partial u }+m{\partial \over
\partial v }\right)+{1\over 2}\,(p\cdot u+mv)^2
\left({\partial \over \partial u}\cdot{\partial \over \partial
u}+{\partial^2 \over \partial
v^2}\right)\right]\varphi=0\,.\label{mfronsdal} \ee
The double
trace constraint on the gauge field becomes \be
\left[\left({\partial \over
\partial u}\cdot{\partial \over \partial u}\right)^2+2\left({\partial \over
\partial u}\cdot{\partial \over \partial u}\right){\partial^2 \over
\partial v^2}
+{\partial^4 \over \partial
v^4}\right]\varphi=0\label{doubletrmass}\,, \ee and the
homogeneity constraint reads \be \left(u\cdot{\partial \over
\partial u}+v{\partial \over \partial v}-s\right)\varphi=0\,.\label{homh} \ee
The equation (\ref{homh}) implies that field $\varphi$ has the
following expansion in the variables $u$ and $v$ \be
\varphi(x,u,v)\,=\,\sum\limits_{r=0}^s{1\over
r!(s-r)!}\,\varphi_{\mu_1\dots\,\mu_r}\,u^{\mu_1}\dots
u^{\mu_r}v^{s-r}\,. \ee Hence the Kaluza-Klein mechanism produces
a tower of totally symmetric tensors of rank going from zero to
$s$. The gauge transformation are \be \delta \varphi\,=\,(p\cdot
u+mv)\, \varepsilon, \label{gtransfS}\ee with \be
\left(u\cdot{\partial \over
\partial u}+v{\partial \over \partial v}-s+1\right)\varepsilon=0 \ee and
the tracelessness of
$\varepsilon$ leads to \be \left({\partial \over
\partial u}\cdot{\partial \over \partial u}+{\partial^2 \over \partial
v^2}\right)\,\varepsilon=0. \ee
Nowadays, this description of massive fields is frequently
referred to as ``St\"{u}ckelberg formulation". General procedures to make connection with
the work of Singh and Hagen \cite{Singh}, by solving the
constraint (\ref{doubletrmass}) and by fixing completely the gauge
transformations (\ref{gtransfS}), were presented in
\cite{Aragone}.

The ``gauge-fixed" equations (\ref{physmassl}) - (\ref{spins})}
become \ba
(p\cdot u+mv)\,\Phi=0,\label{gfm}\\
\left({\partial \over \partial u}\cdot{\partial \over \partial u}
+{\partial^2 \over \partial v^2}\right)\Phi=0,\\
\left(p\cdot{\partial \over\partial u}+
m{\partial \over \partial v}\right) \Phi=0,\\
(p^2+m^2)\,\Phi=0,\label{mspins} \\
\left(u\cdot{\partial \over
\partial u}+
v{\partial \over \partial v}-s+1\right)\Phi =0.\label{ho} \ea

\section{Continuous spin from massive higher-spin}\label{frommhs}

We are now in a position to examine the limit
where the mass goes to zero, the spin to infinity with
their product being fixed. It is clear from equation
(\ref{ho}) that this limit is ill defined on the field $\Phi$.
In order to get a well defined limit, one has to extract an infinite factor
from $\Phi$ and
also to assume a suitable scaling of the scalar $v$.
Let us define the parameter $\mu$ and the variable $\alpha$ by
\be
\mu=sm, \ \alpha={v\over s},
\ee
the precise limit we are interested in is
 when $s$ goes
to infinity with finite $\mu$ and $\alpha$.
Before examining this limit, let us rewrite the massive equations
with the new variables. It will be very convenient to first write the
solution of
equation (\ref{ho})
\be
\left(u\cdot{\partial \over
\partial u}+
\alpha{\partial \over \partial \alpha}-s+1\right)\Phi =0,
\ee
as
\be
\Phi=\alpha^{s-1} \Psi\left({u\over \alpha}\right),
\ee
where we introduced the new field $\Psi$. This is the field
which will remain well defined in the limit.
 If we define $w=u/\alpha$ then
the  equations (\ref{gfm}) - (\ref{mspins}) lead to
\ba
(p\cdot w+\mu)\,\Psi=0\,,\\
\left[{\partial \over \partial w}\cdot{\partial \over \partial w}
+{1 \over s^2}\left((s-1)(s-2)-(2s-3)\left(w\cdot{\partial \over
\partial w}\right)
+\left(w\cdot{\partial \over \partial w}\right)^2\right)\right]\Psi=0\,,\\
\left[p\cdot{\partial \over \partial w}+{\mu \over s^2}
\left(s-1-w\cdot{\partial \over \partial w}\right)\right]\Psi=0\,,\\
\left[p^2+\left({\mu \over s}\right)^2\right]\Psi=0\,, \ea {
using the useful relations \be {\partial\over \partial
u}={1\over\alpha}{\partial\over \partial w}\,,\quad {\partial\over
\partial v}={1\over s}\left(-{w\over \alpha}\cdot {\partial\over \partial
w}+{\partial\over \partial \alpha}\right)\,.\ee
}The limit of infinite spin is now non-singular and we
 get precisely
 the Wigner equations (\ref{cspin1}) -
(\ref{cspin4}). It is now clear that although the field $\Phi$
has an ill defined limit the product $\alpha^{1-s} \Phi$ has a
finite limit which is the continuous spin wave function $\Psi$. We
stress that it was crucial in the above procedure that the
auxiliary variable $v$ grows as $s$ in the limit. In terms of the
Taylor expansion of the original field $\Phi$ this amounts to a
rescaling of the various tensor fields by complicated factors
depending on the spin $s$.

\section{Continuous spin from Fronsdal
equations}\label{fromFronsdal}

In the preceding section, we showed how to obtain the Wigner
equations for the continuous spin particle starting from a gauge
fixed version of the St\"{u}ckelberg formulation. Here, we
perform the contraction directly on the Fronsdal-like equations in
order to find the gauge invariance leading to the continuous spin.
For the spin-$1$ and spin-$2$ particles, the gauge invariance
plays a crucial role in understanding various aspects of the
Maxwell and Einstein theories. For higher-spin fields the gauge
invariance and its deformations are also crucial in discussing
possible interactions. {Generally speaking, gauge invariance
is an important ingredient for deriving covariant field equations from
an action principle.

We solve as before the homogeneity {condition (\ref{homh})
as
\be
\varphi(u,v)=\alpha^s\psi(w),
\ee
where we used the
same change of variables as before. Next, we take the limit of infinite
spin from (\ref{mfronsdal}), rewritten for $\psi$,
and the resulting equation is
the Fronsdal-like continuous spin equation
 \be
\left[p^2-(p\cdot w+\mu)\left(p\cdot {\partial \over
\partial w}\right) +{1\over 2}\,(p\cdot w+\mu)^2\left( {\partial \over
\partial w}\cdot{\partial \over \partial w}+1\right)\right]
\psi=0\,.\label{contFrsond} \ee The double tracelessness
condition (\ref{doubletrmass}) becomes \be \left({\partial \over
\partial w}\cdot{\partial \over
\partial w}+1\right)^2\psi=0\,. \label{dtr}\ee
The gauge transformation parameter can similarly be written as \be
\varepsilon(u,v)=\alpha^{s-1}\epsilon(w)\,. \ee The gauge
transformation becomes \be \delta \psi=(p\cdot w+\mu)\epsilon, \ee
with the gauge parameter $\epsilon$ subject to the trace condition
\be \left( {\partial \over \partial w}\cdot{\partial \over
\partial w}+1\right)\epsilon=0\label{cong}. \ee So the continuous spin
field is now described {\it \`a la} St\"uckelberg. The
relation with the preceding description is obtained by first
fixing the gauge with a de Donder-like condition
\be
\left[p\cdot{\partial \over
\partial w}-{1\over 2}\,(p\cdot w+\mu)\left( {\partial \over \partial
w}\cdot{\partial \over
\partial w}+1\right)\right]\psi=0\,. \ee
The equation of motion
reduces to $p^2=0$ and the residual gauge invariance allows to
impose \be \left( {\partial \over \partial w}\cdot{\partial \over
\partial w}+1\right)\psi=0\,. \ee If we now define
\be\Psi=\delta(p\cdot w+\mu)\,\psi\ee then the wave function
$\Psi$ is gauge invariant and we get back the Wigner equations for
$\Psi$.

A suggestive way of writing the gauge invariant equation is to
decompose $\psi$ as a sum of homogeneous functions $\psi_r$ {
of degree $r$ in $w$, the coefficient of which is a rank-$r$
tensor: \be \psi=\sum_{r=0}^{\infty}\psi_r. \ee The equation
(\ref{contFrsond}) decomposes as} \ba &&\left[p^2-(p\cdot w)
\left(p\cdot{\partial \over
\partial w}\right) +{1\over 2}\,(p\cdot w)^2 \left({\partial \over \partial
w}\cdot {\partial \over\partial w}\right)\right]\psi_r\,=\nonumber \\
&&\quad\quad\quad-\,{1\over 2}\,(p\cdot w)^2\,\psi_{r-2} \nonumber\\
&&\quad\quad\quad-\,\mu\left\{\,(p\cdot w)\,\psi_{r-1}+\left[
(p\cdot w) \left({\partial \over \partial w}\cdot{\partial \over
\partial w}\right)-\left(p\cdot{\partial \over \partial
w}\right)\right]\psi_{r+1}\,\right\} \nonumber\\
&&\quad\quad\quad-\,{\mu^2\over
2}\,\left\{\,\psi_r+\left({\partial \over \partial
w}\cdot{\partial \over
\partial w}\right) \psi_{r+2}\,\right\}\,.\label{coupl} \ea The
{left-hand-side} is just the Fronsdal operator acting on
$\psi_r$ and the right-hand-side contains the couplings of
the rank-$r$ field with the other fields. The constraint
(\ref{dtr}) yields \be \psi_r+2\left({\partial \over
\partial w}\cdot{\partial \over \partial w}\right)
\psi_{r+2}+\left({\partial \over \partial w}\cdot{\partial \over
\partial w}\right)^2 \psi_{r+4}=0. \ee

Similarily to the higher-spin case, the gauge parameter was
constrained by Equation (\ref{cong}). It is possible to remove
this trace constraint by introducing a compensator field
$\chi$ which transforms as \cite{compensator} \be \delta
\chi=\left({\partial \over \partial w}\cdot{\partial \over
\partial w}+1\right)\epsilon\,. \ee The gauge invariant equations
of motion are now given by \be \left[p^2-(p\cdot
w+\mu)\left(p\cdot{\partial \over \partial w}\right) +{1\over
2}(p\cdot w+\mu)^2\left( {\partial \over \partial w}\cdot{\partial
\over\partial w}+1\right)\right] \psi={1\over 2}\,(p\cdot
w+\mu)^3\chi\,, \ee and \be \left({\partial \over
\partial w}\cdot{\partial \over
\partial w}+1\right)^2\psi=\left[ (p\cdot w+\mu)\left({\partial \over
\partial w}\cdot{\partial \over
\partial w}+1\right)+ 2\left(p\cdot{\partial \over \partial
w}\right)\right]\chi\,.
\ee The partial gauge fixing $\chi=0$ gives back the previous
equations (\ref{contFrsond}) and (\ref{dtr}).

\section{Fermionic equations}\label{generalis}

In this section we consider the double-valued continuous spin
representation, formulated with the aid of a $D$-dimensional
spinor.

We start with a massless totally symmetric
spinor-tensor in $D+1$ dimensions\footnote{When $D+1$ is even we can 
start with a Weyl spinor-tensor.}
\be
\varphi^\alpha={1\over
s!}\,\varphi^\alpha_{\mu_1\dots\,\mu_s}\,u^{\mu_1}\dots u^{\mu_s}.
\ee
The Fang-Fronsdal equation reads
\cite{Fang:1978wz}
 \be
\left[\Gamma\cdot p- (u\cdot
p)\left(\Gamma\cdot{\partial\over\partial
u}\right)\right]\varphi=0, \label{ff}\ee
where we used the gamma matrices in $D+1$ dimensions which satisfy
$\{\Gamma^\mu,\Gamma^\nu\}=2\eta^{\mu\nu}$.
Equation (\ref{ff}) is invariant  under the
gauge transformations
\be \delta \varphi=(p\cdot
u)\,\varepsilon\,, \ee
with the spinor-tensor gauge parameter
$\varepsilon$ constrained by the gamma-trace condition
\be
\left(\Gamma\cdot{\partial\over\partial u}\right)\varepsilon=0\,.
\ee The analog of the double trace constraint is now \be
\left(\Gamma\cdot{\partial\over\partial u}\right)\left(
{\partial\over\partial u}\cdot{\partial\over\partial
u}\right)\varphi=0\,. \ee
The homogeneity of $\varphi$ is expressed by
\be
\left(u\cdot{\partial\over\partial
u}-s\right)\varphi=0.
\ee

Next, we compactify on a circle and consider a mode with momentum $m=\mu/s$
along the last direction and then,
 following the same procedure as before, 
 and definig $\psi$ by
 \be
 \varphi=\alpha^s\psi,
 \ee
 we get for $\psi$ in the infinite spin limit,
\be \left[\Gamma\cdot p- (w\cdot p+\mu)\left(\Gamma\cdot
{\partial\over\partial w}+\Gamma^{D+1}\right)\right]\psi=0\,. \label{ffc}\ee
This is the double-valued counterpart of the Fronsdal like equation (\ref{dtr}).
It
 is
invariant under the  St\"uckelberg-like gauge transformations
\be \delta \psi=(w\cdot p+\mu)\,\epsilon\,, \ee with \be
\left(\Gamma\cdot {\partial\over\partial w}+\Gamma^{D+1}\right)\epsilon=0\,.
\ee The double trace constraint becomes \be \left(\Gamma\cdot
{\partial\over\partial
u}+\Gamma^{D+1}\right)\left({\partial\over\partial
u}\cdot{\partial\over\partial u}+1\right)\psi=0. \ee
Notice that the ``square" of the fermionc equation (\ref{ffc})
gives the Fronsdal-like equation (\ref{dtr}).
 
Analogously to the bosonic case, we can now partially fix the gauge
to obtain
\be
\left(\Gamma\cdot
{\partial\over\partial w}+\Gamma^{D+1}\right)\psi=0,\quad \Gamma\cdot p=0
.
\ee
If we define, similarly to the bosonic case,   
the gauge invariant field $\Psi$ by
\be
\Psi=\delta(w\cdot p+\mu)\psi,
\ee
then we obtain the following equations for the double valued continuous spin
field
\ba
(w\cdot p+\mu)\Psi&=&0,\quad (\Gamma\cdot p)\Psi=0,\nonumber\\
 (p\cdot{\partial\over\partial w})\Psi&=&0, \quad
\left(\Gamma\cdot
{\partial\over\partial w}+\Gamma^{D+1}\right)\Psi=0\label{fermion}
\ea
Notice that the massive spinor-tensor that we get by compactification
is non chiral and so in the limit we get a non chiral double valued
continuous spin field. Notice also that a chirality constraint in $D$
dimensions (when $D$ is even)
is not compatible with the above system because the
anticommutator of the last equation with $\Gamma^{D+1}$ gives a constant.
The fourth equation in the system (\ref{fermion})
is absent in the fermionic Wigner equations; it is replaced by its  square
(\ref{cspin2}) and by a chirality
constraint (in $D=4$).
In both cases, the same number of spinorial
components is eliminated. Our formulation is valid for arbitrary dimension $D$.

\section{Exotic representations of the short little
group}\label{mixed}

The representation of the short little group $SO(D-3)$ need
not necessarily be the trivial one {for spacetime dimensions
$D\geqslant 5$ \cite{Brink:2002zx}.

The finite-dimensional representations of the pseudo-orthogonal
groups $SO(p,q)$ correspond to rank-$r$ (gamma)-traceless
(spinor)-tensors with ``mixed symmetries" labeled\footnote{ In
order to discard ``dual" representations of $SO(p,q)$ one may
further restrict the positive integer $c$ to be smaller or equal
to the integer part of ${p+q\over 2}$.} by partition of the
positive integer $r$ into $c$ integer parts $r_1\,$, $r_2\,$,
$\ldots\,$, $r_c$ with \be r_1+r_2+\dots+r_c=r\,,\quad
r_1\geqslant r_2\geqslant \ldots\geqslant r_c>0\,,\ee and $c<p+q$.
Such a partition is denoted by $(r_1\,,r_2\,, \,\ldots\,, r_c)$
and is depicted by the Young diagram 
\be
\begin{picture}(75,70)(-20,2) \multiframe(0,0)(13.5,0){1}(15,10){}
\put(20,0){$r_c$}
\multiframe(0,10.5)(13.5,0){1}(20,10){}\put(25,10.5){$r_{c-1}$}
\multiframe(0,21)(13.5,0){1}(40,20){\ldots}\put(45,25){$\vdots$}
\multiframe(0,41.5)(13.5,0){1}(55,10){} \put(60,41.5){$r_2$}
\multiframe(0,52)(13.5,0){1}(70,10){} \put(75,52){$r_1$}
\end{picture}\label{Young}\ee made of $c$ rows, with the $n$th row
containing $r_n$
boxes. To describe such irreducible representations of the
(pseudo) orthogonal groups, it is convenient to introduce $c$
commuting auxiliary variable (for more details, see {\it e.g.}
Section 3 of \cite{BCIV}).

\subsection{Mixed symmetry gauge fields}

As before we start from the helicity representations to build the
continuous spin representations. During the last years, several
steps have been performed towards a detailed understanding of
mixed symmetry (also called ``exotic") gauge fields in Minkowski
spacetime \cite{Bekaert:2002dt}.

For the sake of simplicity, we will focus on gauge-fixed field
equations generalizing (\ref{physmassl}) - (\ref{spins}) though
one could start from the Labastida equations that generalize the
Fronsdal formulation \cite{Labastida:1987kw}. The wave function
$\Phi(x,u_A)$ is a polynomial in the commuting variables
$u^\mu_A$. The subscripts will run from $0$ to $c$ and be denoted
by capital Latin letters $A\,$, $B\,$, {\it etc}. Proper wave
equations are \ba
p^2\,\Phi&=&0\,,\label{lcone0}\\
(p\cdot u_A)\,\Phi&=&0\,,\label{lcone1}\\
\left(p\cdot {\partial \over \partial u_A}\right) \Phi&=&0\,,\label{lcone2}\\
\left({\partial \over \partial u_A}\cdot{\partial \over \partial
u_B}\right)\Phi&=&0\,,\label{o}\\
\left[\left(u_A\cdot{\partial \over \partial
u_B}\right)-(r_A-1)\,\delta_{AB}\right]\Phi&=&0\,,\quad
(A\leqslant B)\,.\label{gl}
 \ea
Equation (\ref{lcone1}) can be solved as\be\Phi=\delta(u_A\cdot p)
\,\varphi\,, \ee and leads together with Equation (\ref{lcone2})
to the fact that $\varphi$ depends only on the transverse variables
$u_A^i$. Then the last two conditions (\ref{o}) - (\ref{gl})
express the fact that the coefficients of $\varphi(u_A^i)$ belong
to an irreducible tensor representation of the helicity little
group $SO(D-2)$ characterized by the Young diagram \be
\begin{picture}(75,70)(-20,2) \multiframe(0,0)(13.5,0){1}(15,10){}
\put(20,0){$r_c$}
\multiframe(0,10.5)(13.5,0){1}(20,10){}\put(25,10.5){$r_{c-1}$}
\multiframe(0,21)(13.5,0){1}(40,20){\ldots}\put(45,25){$\vdots$}
\multiframe(0,41.5)(13.5,0){1}(55,10){} \put(60,41.5){$r_2$}
\multiframe(0,52)(13.5,0){1}(70,10){} \put(75,52){$r_1$}
\multiframe(0,62.5)(13.5,0){1}(110,10){} \put(115,62.5){$r_0$}
\end{picture}\quad\quad.\label{Youngs}\ee

\subsection{Exotic continuous spin fields}

Performing the Kaluza-Klein compactification leads to a splitting
of the auxiliary variables into $(u_A^\mu,v_A)$. Moreover, one
makes contact with the procedure of Section \ref{frommhs} by
identifying the couple of variables $(u,v)$ with the couple
$(u_0,v_0)$. Let the Latin letters such as $a\,$, $b\,$, $\ldots$
be subscripts running from $1$ to $c$.

The length of the first row in (\ref{Youngs}) may be identified
with the ``spin" $s=r_0\,$. The infinite spin limit $s\rightarrow
\infty$ of the massive equations with $A=B=0$ leads to the Wigner
equations (\ref{cspin1}) - (\ref{cspin4}), exactly as in Section
\ref{frommhs}. The novelty is that they are supplemented by new
equations.

To start with, one should look at the equation
\be
\left[\left(u_0\cdot{\partial \over \partial
u_b}\right)+v_0\,{\partial \over \partial v_b}\right]\Phi=0\ee
which implies that
\be {\partial \Psi\over
\partial v_a}=-{1\over s}\left(w\cdot{\partial\Psi \over
\partial u_a}\right)\,,\label{condpartv}\ee
where we recall that $\Phi=\alpha^{s-1}\Psi$ and $v_0=s\alpha$.
Thus, the wave function does not depend on $v_a$ in the infinite spin limit.
 The condition (\ref{condpartv}) allows to derive the last
equations \ba (p\cdot  u_a)\,
\Psi&=&0\,,\label{new2.5}\\\left(p\cdot {\partial \over \partial
u_a}\right)
\Psi&=&0\,,\label{new3}\\
\left({\partial \over \partial w}\cdot {\partial \over \partial
u_a}\right) \Psi&=&0\,\\
\left({\partial \over \partial u_a}\cdot{\partial \over \partial
u_b}\right)\Psi&=&0\label{new1}\,,\\
\left[\left(u_a\cdot{\partial \over \partial
u_b}\right)-(r_a-1)\,\delta_{ab}\right]\Psi&=&0\,,\quad
(a\leqslant b)\,.\label{new2}\label{new4}\ea

As before, Equations (\ref{cspin1}) and (\ref{new2.5}) can be
solved as\be\Psi=\delta(w\cdot p+\mu)\,\delta(u_a\cdot p)
\,\psi\,. \ee Then Equations (\ref{cspin3}) and (\ref{new3}) imply
in turn that $\psi$ depends only on the transverse variables $w^i$
and $u_a^i$. Equation (\ref{new4}) further eliminates one direction
of the variables $u_a$. Eventually, Equations (\ref{new1}) -
(\ref{new2}) impose that the physical components are in an
irreducible representation of the short little group $SO(D-3)$
depicted by the Young diagram (\ref{Youngs}). This result is in
complete agreement with the group-theoretical analysis of
\cite{Brink:2002zx}.

To summarize the single valued exotic case, one may say that the
Wigner equations should be supplemented with the set  (\ref{new2.5}) - (\ref{new2}). In the
double valued exotic case, one should further add Equations
(\ref{fermion}) and the set
\be
\left(\Gamma\cdot{\partial\over\partial u_a}\right)\Psi=0\,.\ee
which expresses the gamma tracelessness for each index.

\section{Conclusions and perspectives}

We obtained covariant wave equations for the continuous spin field
in various forms as an infinite spin limit of massive higher-spin equations.
The first ones are identical to Wigner equations and were obtained from the gauge
fixed equations. The second ones arise from the Fronsdal equations
and exhibit   gauge symmetries with constrained or
unconstrained gauge parameters.
A natural question arises: Is it
possible to derive some of these equations from an action
principle? In this respect, we mention that the limit of the ``Einstein tensor" is
singular even with the field redefinitions we performed.
Thus the possibility of formulating the action principle remains open.

Higher-spin fields are known to propagate consistently in constant
curvature backgrounds, therefore the infinite spin limit suggests
to look  for the flat space limit
of some $(A)dS_D$ representations that would lead to the
continuous spin representations.
The mass-shell condition for
totally symmetric representations of the anti de Sitter group
$SO(D-1,2)$ is \cite{Metsaev} \be \Big[D^2-m^2-{1\over R^2
}\,\Big(s^2+(D-6)s+2(3-D)\Big)\Big]\Phi=0\,,\label{mshellAdS}\ee
where $D$ is the covariant derivative and $R$ is the radius of
$AdS_D$. In order for the infinite spin limit of (\ref{mshellAdS})
to be non-singular it appears to be indeed necessary to require that the
radius $R$ goes to infinity at least as fast as $s$. This argument
also holds for any massive representation of the $AdS_D$ isometry
group.

Looking at the continuous spin equation expanded in the homogeneity degree,
 the parameter $\mu$ plays the role of a
coupling constant of bilinear interactions between fields of
different degrees. These coupling terms are responsible for the fact
that the representation is irreducible. In the decoupling limit
$\mu\rightarrow 0$ the single (or double) valued continuous spin
representation, as we showed in Section 2.2,
indeed decomposes as an infinite sum of helicity
representations for all integer (or half-odd) spins. Therefore
this limit provides a natural mechanism to generate an infinite
tower of massless higher-spins. The latter seems to be a proper
starting point to switch on interactions (see {\it e.g.}
\cite{Vasiliev:2004qz,BCIV}) so it would be interesting to try
introducing self-interactions for the continuous spin field
itself. \footnote{ Even in flat spacetime, the fact that the
parameter $\mu$ has mass dimension is promising for writing
consistent interaction vertices, since $\mu$ might play a role
analogue to the inverse $AdS$-radius $1/R$ in Vasiliev's
construction.}

At first sight, second quantisation 
of the continuous spin field seems to lead either to non-locality or to a
breakdown of causality \cite{Abbott:1976bb}. Wigner himself argued
against the continuous spin particles because they should lead to
an infinite heat capacity of the vacuum, essentially because the
number of polarizations ({\it i.e.} the spin in $D=4$) is infinite
\cite{Wigner:1963in}.
It would be very satisfactory
if the infinite spin limit could explain -- if not resolve -- the
elusive properties of the continuous spin representation. Since
the original higher-spin massive field is well behaved, the subtle
infinite spin limit should be at the origin of these strange
properties and finding a proper limit might
regularise some of those unconventional characteristics.

\acknowledgments

X.B. thanks the I.H.\'E.S. for hospitality. He is supported in part by the European Research Training Network
contract 005104 ``ForcesUniverse".

\end{document}